\documentclass[11pt,a4paper]{article}
\pdfoutput=1
\usepackage[hyperref]{nlp4MusA}
\usepackage{times}
\usepackage{latexsym}
\usepackage{amsmath} 
\usepackage{makecell, booktabs} 
\usepackage{graphicx} 
\usepackage{xcolor}

\usepackage{url}

\aclfinalcopy 
\def\aclpaperid{10} 

\title{Improving Real-time Score Following in Opera \\ by Combining Music with Lyrics Tracking}

\author{Charles Brazier$^1$ \hspace{0.5cm} Gerhard Widmer$^{1,2}$ \\ 
$^1$Institute of Computational Perception, Johannes Kepler University Linz, Austria\\
$^2$LIT AI Lab, Linz Institute of Technology, Austria\\
\tt{firstname.lastname}@jku.at}

\date{}

\begin{document}
\maketitle
\begin{abstract}
Fully automatic opera tracking is challenging because of the acoustic complexity of the genre, combining musical and linguistic information (singing, speech) in complex ways. In this paper, we propose a new pipeline for complete opera tracking. The pipeline is based on two trackers. A music tracker that has proven to be effective at tracking orchestral parts, will lead the tracking process. In addition, a lyrics tracker, that has recently been shown to reliably track the lyrics of opera songs, will correct the music tracker when tracking parts that have a text dominance over the music. We will demonstrate the efficiency of this method on the opera \textit{Don Giovanni}, showing that this technique helps improving accuracy and robustness of a complete opera tracker.
\end{abstract}

\section{Introduction and Contribution}

Score following aims at aligning classical music performances with their corresponding scores (sheet music), in order to assign a score position at each time step in the performance. There has been constant progress in this domain, starting with the tracking of monophonic melodies in \cite{dannenberg1984line}, all the way to recent systems that can follow, under real conditions, complex orchestral works \cite{arzt2015real} in a completely autonomous process. This has led to the development of new applications such as automatic page-turning for pianists \cite{arzt2008automatic}, live performance visualization \cite{lartillot2020real}, or score viewing and automatic contextualization in orchestra concerts \cite{prockup2013orchestral,arzt2015artificial} to enrich the viewers' experience.

Tracking live opera performances would become an essential tool for all future opera halls, supporting functionalities like fully automatic subtitles display, or automatic camera control and video editing for live streaming services. However, and compared to previous existing works, operas are more challenging to track, due to the setup with a complete orchestra and singers that act and sing on stage, one or several at a time, for several hours, with various noises, acting breaks, intermittent applause, musical (sometimes improvised) interludes, etc.

First attempts at opera tracking \cite{brazier2020towards, brazier2020addressing} use an \textit{On-Line Dynamic Time Warping (OLTW)} algorithm \cite{dixon2005line} to align complete performances with a \textit{reference performance} (some other recording of the work in question) that has been aligned to the score beforehand and serves as a proxy to the score. This audio-to-audio alignment strategy is an elegant way to circumvent the unavailability of complete opera score files in symbolic format. Also, using a real recording is advantageous because the sounds in the reference are much more realistic and similar to what is to be expected in the real performance than anything one could synthesize from a score. \citet{brazier2020towards} combine alignment with three audio event detectors for music, speech/singing voice, and applause, which halt the tracking process during long silences, applause, or interlude passages that can occur in between the parts. \citet{brazier2020addressing} further improve tracking accuracy by using two trackers working in parallel, one using audio features tuned on orchestral music \cite{gadermaier2019study}, the other using features tuned on the \textit{recitative} subset of one opera performance.

In this work, we propose to exploit an additional source of information: the \textit{lyrics} sung or spoken in the audio recordings. We do not assume the written lyrics to be available in textual form. Rather, the idea is to train an acoustic phoneme recognition model that extracts phoneme sequence estimates both from the reference (off-line) and the live performance (on-line), and to align these in real time, giving us a real-time lyrics tracking algorithm. More specifically, the acoustic model will predict, for each audio frame, a probability vector over a set of phonemes. For each part (aria, recitative, etc.) in the score, we assign a \textit{voice on music ratio} value, calculated on the reference recording with the help of dedicated music/speech audio classifiers. The music tracker leads the alignment process. As soon as the score position corresponds to a voice-dominant part in the score, the lyrics tracker starts and we rely on its score position. When the score position reaches a music-dominant part, the lyrics tracker is stopped and the music tracker alone is used.

Acoustic model and lyrics tracker have already been presented in a recent publication \cite{brazier2021online}, but only evaluated on selected text-heavy \textit{recitativo} passages. Here we demonstrate, for the first time, the benefit of combining lyrics with music tracking in an automated way.

\section{Data Description}
\label{sec:data}

Score followers are evaluated by computing their alignment accuracy on audio performances that have been manually annotated to the corresponding score. As no such dataset exists for opera, we had to create our own. The dataset focuses on the opera \textit{Don Giovanni} by W.A.Mozart. As the \textit{reference}, serving as a proxy to the score, we selected a commercial CD recording conducted by Herbert von Karajan in 1985. As \textit{target performances} that we want to align to the score in real time, we use two full live performances, with different casts and stagings, that have been recently recorded at and by the \textit{Vienna State Opera}, one conducted by \'Adam Fischer in 2018, the other by Antonello Manacorda in 2019. There are two parts in the reference that are not played in the two live recordings. For this study, we decided to remove these to align performances that follow the same score structure. Compared to the reference, the live performances contain applause, breaks, and interludes that can appear between parts. The dataset details are given in Table \ref{tab:dataset}.

\begin{table}[t]
\begin{center}
\scalebox{0.8}{
\begin{tabular}{@{}llccl@{}}
\toprule
\textbf{Conductor} & \textbf{Place} & \textbf{Year} & \textbf{Duration} & \textbf{Role} \\
\midrule
H.v. Karajan & Berlin & 1985 & 2:57:53 & Reference \\
\'A. Fischer & Vienna & 2018 & 3:12:54 & Target \\
A. Manacorda & Vienna & 2019 & 3:07:09 & Target \\
\bottomrule
\end{tabular}}
\end{center}
\caption{Dataset used in this study.}
\label{tab:dataset}
\end{table}

For each performance in the dataset, we manually affixed 5,304 bar annotations, 2,866 for the first act and 2,438 for the second, corresponding to the total number of bars present in the 500-pages score book. The annotations in the reference performance permit to link the complete performance to the score book. The annotations in the target performances serve for evaluating the alignment accuracy of our tracker. Thus, our dataset comprises more than 9~hours of opera recordings played and sung in real conditions by different orchestras and singers and recorded with different recording setups. It contains around 16,000 manual bar-level annotations assigned to the 530 pages score book, which is available online thanks to the \textit{Mozarteum Foundation Salzburg}\footnote{https://dme.mozarteum.at/DME/nma/}. Precisely, annotating these 9 hours of music took about 300 hours of work.

\section{Real-Time Opera Trackers}
\label{sec:trackers}

Operas are complex works that combine music, singing, and speech in complex ways. Most of the time, the piece is led by the music, with singers singing on top of the orchestra. However, operas also include passages, such as \textit{recitativo} sections, where the dominant signal is the lyrics spoken or sung by the singers, with a sparse musical accompaniment that is played differently across performances (e.g., arpeggiated chords, not aligned to the lyrics, partly improvised, and played by different instruments). To tackle this, we propose to alternate between two trackers, one focusing on the music information and the other on the lyrics information. We first describe our \textit{music tracker} that serves as a baseline in this study and that leads the tracking process. We then describe our \textit{lyrics tracker}, and then propose one simple way of combining them for achieving a better global tracking accuracy. This combination strategy will be experimentally verified in the next chapter.

\subsection{Music Tracker}
\label{sec:music_tracker}

The \textit{music tracker} is based on an adaptive version of the On-Line Time Warping (OLTW) algorithm \cite{dixon2005line} that has been successfully used in orchestra \cite{arzt2008automatic} and also in opera tracking \cite{brazier2020towards, brazier2020addressing}. The OLTW algorithm updates an accumulated cost vector that has the length of the reference feature sequence, where the index of its minimal value corresponds to the score position given by the algorithm. Per audio frame, it receives as input a feature vector of 100 MFCCs (120 MFCCs are calculated from the audio sampled at 44.1~kHz, but the first 20 are discarded \cite{gadermaier2019study}), computed with a window size of 20~ms, and a hop size of 10~ms. The features of the reference audio are computed beforehand, while those of the target performance are computed in real-time. For each new incoming target feature, we compute the cosine distance between the feature and an interval of reference features of length $c$, centered around the expected score position (in practice $c$ is fixed to 4000, corresponding to a context of 40~seconds of audio). Then, considering the previous score position $sp$, the previously accumulated cost vector $D_{prev}$, and the current distance vector $d$, we compute the value of the new accumulated cost vector $D$ by first initializing its values by $+ \infty$, and then applying the following recursive formula:

$\forall i\in \left[sp-c/2:sp_{j-1}+c/2\right],$
\begin{equation}
D[i] = d[i-(sp-c/2)] + \min \begin{cases}
        D_{prev}[i-1]\\
        D_{prev}[i]\\
        D[i-1]\\
    \end{cases}
\label{eqn:OLTW}
\end{equation}

To compare costs in $D$ among themselves and not favor shorter paths over longer ones, we normalize them by dividing all values by their distance from the initial score position (i.e. by the sum of their index in the accumulated vector and an incremental counter representing the number of iterations since the beginning of the tracking).

Our target performances are performed under real conditions and thus include applause, breaks, or interludes that can be played in between the parts. We make use of the applause, music, and speech detectors detailed in \cite{brazier2020addressing} to halt the tracking process when detected.

\subsection{Lyrics Tracker}
\label{sec:lyrics_tracker}

The \textit{lyrics tracker} makes use of an on-line audio-to-lyrics alignment method that has been shown to robustly track the lyrics of different languages, in the genre of opera \cite{brazier2021online}. The tracker is composed of an acoustic model that generates, in real-time, \textit{posteriograms} representing the frame-wise probability distribution over a set of predefined phonemes through time. Then, it employs the same OLTW algorithm described in Section \ref{sec:music_tracker}, but in this case, aligning the posteriogram of the reference performance generated beforehand, and the posteriogram of the target performance generated online. This obviates the need for a text-to-phoneme tool to translate the written-out lyrics, as well as a manual alignment of the lyrics to the reference performance. It works without having the lyrics themselves and can track a language other than the language(s) the acoustic model was trained on, as shown in \cite{brazier2021online}.

The \textit{acoustic model} is the core element of our lyrics tracker; its role is to estimate in real-time a posteriogram matrix from the audio recording. Its architecture is the CP-ResNet \cite{koutini2019receptive}, composed of convolutional layers with residual connections between layers, and has a receptive field of 57 frames in the input feature sequence centered around its time position, fixing the delay of the model to 28 frames. The model takes as input 80~MFCCs that are extracted from an audio window of 20~ms, sampled at 16~kHz, with a hop size of 10~ms; it outputs a vector every 40~ms. The output vector is of length 60, representing the classes of the 57~different phonemes that are included in the multilingual DALI dataset \cite{meseguer2018dali} used to train the model. The dataset collects 275~hours of Western musical genres with lyrics annotations at the sentence, word, or note level, and includes English, German, French, Spanish and Italian languages. The phoneme representation permits to train a single model on different languages \cite{vaglio2020multilingual}. The output vector also adds the space token, the instrumental token, and the blank token, essential to a Connectionist Temporal Classification (CTC) training \cite{graves2006connectionist} (the blank class will be ignored when applying Equation \ref{eqn:OLTW}).

\subsection{M\&L Tracker: Combining Music and Lyrics Trackers}
\label{sec:ML_tracker}

To exploit the complementarity between the two previously described trackers, we first classify each part of the opera in two classes (in the given reference performance): parts dominated by the music and parts dominated by the voice. To do so, we use the structure detailed in the Table of Contents of the Opera\footnote{\url{dme.mozarteum.at/DME/nma/nma_toc.php?vsep=68}}, and consider each title as an individual part. For each part, we use the music and voice detectors (already used to halt the tracking process in between parts, as mentioned in Section \ref{sec:music_tracker} above) to calculate a \textit{voice over music ratio} that is given by the percentage of voice along the part divided by the percentage of music. Thus, an instrumental part will have a ratio close to 0, whereas a part that contains more voice than music will have a ratio higher than 1.

The combination of the two proposed tracking models is delicate because they both work at a different pace (10ms for the music tracker, and 40ms for the lyrics tracker), the lyrics tracker has a delay of 280ms in its output due to its receptive field, and neither of them is able to track accurately full opera performances. More precisely, the music tracker is inaccurate when an improvised accompaniment is played during a part led by the lyrics, and the lyrics tracker is entirely lost during instrumental parts. Our approach is to use the music tracker continuously, along with the complete target performance. When the score position given by the music tracker corresponds to a part in the score that, according to our estimated voice/music ratio, is dominated by voice(s), we initialize the accumulated cost vector of the lyrics tracker by values of $+ \infty$ everywhere, and a value of 0 at the score position given by the music tracker. We then use separately music and lyrics trackers but we rely only on the score position given by the lyrics tracker. As soon as the score position given by the lyrics tracker corresponds to a part in the score dominated by music, we stop the lyrics tracker and rely on the position given by the music tracker.

\section{Experiments and Discussion}
\label{sec:experiments}

For our experiments, we compare three different tracking models. The first, \textit{music}, reproduces the work in \cite{brazier2020towards} and uses the music tracker only (including acoustic event detectors to deal with interludes and other unexpected events such as applause and acting pauses). The second one, \textit{musicP}, is the state-of-the-art opera tracker \cite{brazier2020addressing}; it uses two music trackers in parallel, one using the features detailed in Section~\ref{sec:music_tracker}, the other using optimized audio features that have been tuned on the recitative subset of the Fischer performance. Finally, the third tracker \textit{M\&L} is the contribution of this paper. The systems are evaluated by their alignment accuracies \cite{cont2007evaluation}. We report the mean error in ms, as well as the proportions of bar boundaries (which reflect the precision of our ground truth annotations) that are detected with an error less than 1, 2, and 5 seconds. The results are given in Table~\ref{tab:results}.

\begin{table}[t]
\begin{center}
\scalebox{0.8}{ 
\begin{tabular}{@{}llrrrr@{}}
\toprule
\textbf{Conductor} & \textbf{Tracker} & \textbf{Mean} & $\mathbf{\leq 1s}$ & $\mathbf{\leq 2s}$ & $\mathbf{\leq 5s}$\\
\midrule
\textbf{Fischer}   & music  & 811ms & 91.8\% & 95.0\% & 97.3\% \\
                   & musicP & 373ms & 93.4\% & 96.8\% & 99.0\% \\
                   & \textbf{M\&L}   & \textbf{335ms} & \textbf{94.1\%} & \textbf{97.3\%} & \textbf{99.2\%} \\
\midrule
\textbf{Manacorda} & music  & 561ms & 90.1\% & 94.5\% & 97.9\% \\
                   & musicP & 547ms & 90.3\% & 94.7\% & 98.0\% \\
                   & \textbf{M\&L}   & \textbf{410ms} & \textbf{91.6\%} & \textbf{95.9\%} & \textbf{99.0\%} \\
\bottomrule
\end{tabular}}
\end{center}
\caption{Tracking error of three trackers: music \cite{brazier2020towards}, musicP \cite{brazier2020addressing}, and Music and Lyrics (M\&L).}
\label{tab:results}
\end{table}

For both live target performances, the proposed \textit{music \& lyrics} tracker achieves the best accuracy, beating the \textit{music} tracker, and also the \textit{musicP} tracker whose features were tuned on the Fischer performance. The accuracy improvement on Fischer is relatively small, but no fine-tuning on features is done in our proposal. The improvements on Manacorda are more substantial, dropping the mean error to 410~ms and increasing all the 3 percentages by at least one point.

We tried to take into account the delay of the lyrics tracker, in adding an offset to the score position given by the tracker, but the best results were achieved in ignoring this delay.

\section{Conclusion}
\label{sec:conclusion}

We have presented a new state-of-the-art method for tracking full-length opera performances. The method makes use of an acoustic model that estimates the sung lyrics (phoneme probability vectors) over time. The final model combines lyrics and music information (without requiring the written lyrics as input) via two specific trackers. The combination helps to improve the tracking accuracy of the performance.

The proposed method requires a part segmentation of the reference performance. The beginnings and ends of each part are directly given by the manual bar annotations, useful to also handle structural mismatches in opera \cite{brazier2021handling}. However, we plan to emancipate ourselves from the manual annotations with the development of a method that fully autonomously segments a piece.

\section*{Acknowledgments}

The research is supported by the European Union under the EU's Horizon 2020 research and innovation programme, Marie Sk\l{}odowska-Curie grant agreement No.~765068
(``MIP-Frontiers").
The LIT AI Lab is supported by the Federal State of Upper Austria.

\bibliography{nlp4MusA}
\bibliographystyle{nlp4MusA_natbib}

\end{document}